\documentstyle[11pt,aaspp4]{article}

\def\degr{\hbox{$^\circ$}}

\slugcomment{to appear in Astronomical Journal}
\lefthead{Plana et al.}
\righthead{The kinematics of the warm gas in Hickson compact group of
galaxies HCG 90}

\begin{document}

\title{The kinematics of the warm gas in the interacting 
Hickson compact group of
galaxies HCG 90}

\author{H. Plana and  C. Mendes de Oliveira \altaffilmark{1}}
\affil{Instituto Astron\^omico e Geof\'\i sico (IAG), Av Miguel St\'{e}fano 
4200 CEP: 04301-904 S\~{a}o Paulo Brazil} 

\and

\author{P.Amram and J. Boulesteix}
\affil{IGRAP, Observatoire de Marseille, 2 Place Le Verrier, F-13248 Marseille 
Cedex 
04, 
France}

\altaffiltext{1}{Present address: Universitaets-Sternwarte, 
Ludwig-Maximilians-Universitaet 
Scheinerstrasse 1, 81679 Muenchen, Germany.}

\begin{abstract}

 We present kinematic observations of H$\alpha$ emission for two
early-type galaxies and one disk system, members of the Hickson compact
group 90 (HCG 90) obtained with a scanning Fabry-Perot interferometer
and samplings of 16 $km$ $s^{-1}$ and 1$\arcsec$.  Mapping of the gas
kinematics was possible to $\sim$ 2 r$_{eff}$ for the disk galaxy N7174
and to $\sim$ 1.3 r$_{eff}$ and $\sim$ 1.7 r$_{eff}$ for the early-type
galaxies N7176 and N7173 respectively. Evidence for ongoing
interaction was found in the properties of the warm gas of the three
galaxies, some of which do not have stellar counterparts.

   The system H90bd (N7176-N7174) which was previously suspected to be
an optical double may in reality be a system in interaction.  In the
region where the galaxies spatially overlap in projection (their continuum 
centers are only 25$\arcsec$ apart), the gas profiles are separated in
velocity space by $\sim$ 50--100 km s$^{-1}$. The gas component of the
early-type galaxy is highly concentrated in the region closest to the
irregular galaxy, where the interaction between the two galaxies may be
taking place.  The velocity fields of these galaxies are disturbed,
most probably due to the ongoing interaction; they indicate that
the galaxies are in pro-grade orbits, which is a favourable
condition for merging.

    H90c, morphologically classified as an elliptical galaxy, has a
disk of ionized gas which rotates around an axis oriented 60$\degr$ with
respect to the stellar rotation axis.  This is a strong evidence that
the gas has an external origin.  As is also the case for H90b, this
galaxy may be a true S0 that was misclassified as an elliptical
galaxy.

  We suggest the following evolutionary scenario for the system.  H90d
is the warm gas reservoir of the group in process of fueling H90b with
gas.  H90c and d have experienced past interaction with gas
exchange. The gas acquired by H90c has already settled and relaxed but
the effects of the interaction can still be visible in the morphology
of the two galaxies and their stellar kinematics.  This process will
possibly result in a major merger.

\end{abstract}


\section{Introduction}

   Hickson compact groups of galaxies are small systems of three to
 seven galaxies with projected separations on the sky of a few galactic
 diameters and velocity dispersions comparable to the internal stellar
 motions of the group galaxies.  Although several Hickson groups have a
 high projected galaxy density and seem to form a physical group, the
 hypothesis that a close system in projection is also compact in 3-D is
 only confirmed when the member galaxies are found to be in interaction
 with one another.  If so, the low relative velocity dispersion of
 galaxies in the group as well as the small distances between them will
 favor the formation of merger remnants (Barnes 1989).

   Several photometric and spectroscopic indicators of galaxy
 interactions have been identified in the literature such as shells,
 ripples, isophote twists, double nuclei, kinematically decoupled cores
 in ellipticals (Bertola et al. 1988) and disturbed rotational velocity
 curves of disks (Rubin et al.  1991).  Other useful indicators of
 galaxy interactions can be found in the cold and warm gaseous
 component of a galaxy. Cold gas is very important in identifying
 late-type but not early-type interacting galaxies, since these are
 mostly devoid of HI (van Gorkom 1991, van Gorkom 1997). Warm gas, in
 contrast, is present in both early and late-type galaxies and can be
 used as a tracer of recent or on-going interactions.  Although the
 warm gas component usually contributes only a small fraction of the
 total mass of a galaxy, it responds very quickly to gravitational
 perturbations and should therefore allow a detailed study of the
 recent history of interactions/accretions of the system.

   We have begun a program to study the warm gas properties and
 kinematics of galaxies in a subsample of Hickson compact groups with
 the main aim of identifying interaction and merging signs among the
 member galaxies and in this way attempt to determine the evolutionary
stage of the compact group. Paper I (Mendes de Oliveira et al. 1998)
 was dedicated to the compact group HCG 16, a system in an advanced
 evolutionary stage containing four late-type galaxies, where three of
 the members were shown to be merger remnants from the study of their
 gas properties.

   The object of the present study is HCG90 (also called M59, N7171,
LGG 450, Klemola 34), a group at a distance of 33.15 
h$^{-1}$ Mpc (Faber et al. 1989, $h$ is the dimensionless Hubble
constant $H_o$/100 km sec$^{-1}$ Mpc$^{-1}$) containing two late-type
galaxies, NGC 7172 (H90a) and NGC 7174 (H90d), and two early-type
galaxies, NGC 7176 (H90b) and NGC 7173 (H90c).  This group was
originally cataloged by Hickson (1982).  The mean systemic velocity and
velocity dispersion of the quartet were determined to be 2643 km
s$^{-1}$ and 100 km s$^{-1}$ respectively (Hickson et al. 1992).  The
spectroscopy survey of a circular region with diameter d = 1$\degr$
($\sim$ 0.6 h$^{-1}$ Mpc) centered on the group confirmed the result of Ramella
et al. (1994) that H90 is a compact system surrounded by a loose group
(de Carvalho et al 1997).  Five new members were identified by this
study.  The velocity dispersion of this enlarged system of nine
galaxies is 166 km s$^{-1}$.  Yet more new members were identified when
a larger area around HCG 90 was surveyed (1.5 $\times$ 1.5 degrees) by
Zabludoff and Mulchaey (1998). A total of 16 galaxies were found at the
redshift of the group, with a velocity dispersion of 190 km s$^{-1}$.
 A general trend of constant velocity dispersion with increasing radius
of the group was detected for a combined sample of nine x-ray detected
groups including HCG 90 suggesting that the groups may be embedded in a
common dark matter halo (Zabludoff and Mulchaey 1998).

   The four galaxies in the core of group HCG 90 are infrared emitters,
detected both at 60 and 100 $\mu$m (Allam et al. 1996, for H90b and 
H90c only upper limits for the fluxes are available).  Only H90a is a
radio emitter in 6 and 20 cm (Menon 1995).  Neutral hydrogen was
detected for H90a and not for the other members of the group
(Oosterloo \& Iovino 1997).  CO was detected in all galaxies (Boselli
et al.  1996,  Verdes Montenegro et al. 1998).  The spectroscopic
study of Coziol et al. (1997) showed that H90a and d have active
cores.  H90d is classified as an AGN and H90a is a Seyfert 2
galaxy.  A halo of difuse x-ray  emission was detected around the
group to a radius of 135 h$^{-1}$ kpc (Mulchaey and Zabludoff 1998). The
emission is not centered on any galaxy of the group.

   A study of the stellar kinematics of three of the member galaxies (H
90b, c and d) was done by Longo et al. 1994 (hereafter referred to as
L94).  Their study was based on spectra obtained using long-slit
observations along two position angles of the galaxies (major axis and
on one other position angle).  Those data are used in Section 4.1 of
this paper for a comparison with the properties of the gas kinematics
of the galaxies.

    In the present paper we present Fabry-Perot data of three galaxies
of HCG 90: two early-type galaxies (H90b and c) and one late-type
spiral galaxy (H90d).  The observations and data reduction are
presented in section 2. In section 3 we present a general description
of each object with comments on the velocity fields (VFs),
monochromatic maps and the velocity curves (VCs). The discussion is in
section 4. It includes a comparison between the warm gas and the
stellar kinematics of the galaxies.

\section{Observations and data reduction}

   A two-hour exposure on galaxies H90b, c and d of HCG 90 has been
obtained using the 3.6m ESO telescope in La Silla (Chile) on August
28th, 1995. The instrument CIGALE (CIn\'{e}matique des GALaxiEs,
Boulesteix et al. 1983) attached to the Cassegrain focus was used.
CIGALE is composed of a focal reducer (bringing the original f/8 focal
ratio of the Cassegrain focus to f/2), a scanning Fabry-Perot, a narrow
band interference filter and an Image Photon Counting System (IPCS)
detector.  The IPCS, with a time sampling of 1/50 s and zero readout
noise makes it possible to scan the interferometer rapidly (typically 5
s per channel) avoiding sky transparency, airmass and seeing variation
problems during the exposures.  The basic principles of this instrument
were described in Amram et al. (1991).  The journal of the observations
are given in Table 1.  Table 2 lists the parameters for the three
observed galaxies.  The spatial and spectral sampling are 0.91$\arcsec$
and 16 km s$^{-1}$ respectively.

  Reduction of the data cubes were performed using the CIGALE/ADHOC
software (Boulesteix 1993).  The data reduction procedure has been
extensively described in Amram et al. (1996) and references
therein.

  Wavelength calibrations were obtained by scanning the narrow Ne 6599
\AA\ line under the same conditions as the observations.  The relative
velocities with respect to the systemic velocity are very accurate,
with an error of a fraction of a channel width (${\rm <3 \, km
s^{-1}}$) over the whole field.

    The signal measured along the scanning sequence was separated into
two parts: (1) an almost constant level produced by the continuum light
in a 10 \AA ~passband around H$\alpha $ (continuum map), and (2) a
varying part produced by the H$\alpha $ line (monochromatic map).  The
continuum level was taken to be the mean of the three faintest
channels, to avoid channel noise effects.  The monochromatic map was
obtained by integrating the monochromatic profile in each pixel.  The
velocity sampling was 16 $km$ $s^{-1}$.  The monochromatic maps had
one-pixel resolution in the center of the galaxies. Spectral profiles
were binned in the outer parts (to 5 $\times$ 5 pixels) in order to
increase the signal-to-noise ratio.  When multiple components were
visually present, the lines were decomposed into multiple Gaussian
components.  OH night sky lines passing through the filters were
subtracted by determining the shapes and intensities of the lines away
from the galaxies (Laval et al. 1987).

  A rough flux calibration of the monochromatic images  was attempted
by adjusting the flux levels to those of the calibrated image of the
Cartwheel galaxy, obtained in the same run (see details of how the
Cartwheel galaxy image was calibrated in Amram et al. 1998). H$\alpha$
profiles for the HCG 90 galaxies were measured to a minimum flux density
between 0.2 $\times$ 10$^{-17}$ erg s$^{-1}$cm$^{-2}$ arcsec$^{-2}$ and
3 $\times$ 10$^{-17}$ erg s$^{-1}$cm$^{-2}$ arcsec$^{-2}$
(corresponding to a S/N between three and five).

%
%Table 1 Journal of observations
%

\placetable{tbl-1}

%
%Table 2 Physical parameters
%

\placetable{tbl-2}

\section{Results}

   Fig. 1 presents a contour plot of the continuum map of the H90
group.  The total field of view is 3.5$\arcmin$ $\times$ 3.5$\arcmin$.
The continuum map was made using a spatial square smoothing box of
3$\times$3 pixels.  Figs. 2 to 8 show the monochromatic images, the
velocity fields (VFs) of $H\alpha$, the velocity curves (VCs) and a
plot of the variation of the kinematic PA with radius.  The
line-of-sight velocities plotted have not been adjusted by the
cosmological correction ({\it~1~+~z~}).

   We can directly compare the line-of-sight velocity curves
of the three galaxies studied here, since by chance their
inclinations are very similar ($\sim$ 55$\degr$).  Therefore,  no
deprojection of the velocities was done, for the plots in Figs. 3, 6
and 8, although the maximum velocities listed in Table 2 do
include the deprojection as do also the determinations of the masses.

 Detailed discussion on the method used to obtain the systemic
velocity, center, position angle (PA) of the major axis and inclination
of the galaxy are given in Amram et al. (1996).

\subsection{H90c (NGC 7173)}

H90c has been classified as an E0 galaxy by Hickson (1993) and as an
E+pec galaxy in the RC3 catalogue.  Its low ellipticity ($\epsilon$ =
0.26) makes it difficult to determine the PA of its major axis.  In the
inner part (out to 10") it is $\sim$ 60$\degr$ and in the
outskirts of the galaxy it reaches 130$\degr$ (all PA measurements are
made from north to east).

   Figs. 2a and 2b present respectively the monochromatic image and the
velocity field.  Emission is measured out to a radius of $\sim$
25$\arcsec$, which represents 1.7 $r_{eff}$ or 4 h$^{-1}$ kpc (using $r_{eff}$
given by L94). We can see, on Fig.2a, a small excess of gas in a region
located to the north at 6.6" from the center.

  The velocity field for H90c presents a peculiarity in the SE, where
some of the iso-velocities with signal-to-noise S/N $\sim$ 3
(velocities 2800 km s$^{-1}$ and 2820 km s$^{-1}$) are inconsistent
with the rest of the map.  In the center of the galaxy, the S/N of the
data is typically 30.

The variation of the PA of the kinematic major axis with radius for
H90c (not presented here) is small. From the center to 22$\arcsec$ the
PA varies continuously from 20$^o$ to 40$^o$.  The line-of-sight
velocity diagram (hereafter referred to as VC for velocity curve)
presented in Fig. 3a, has been built with a PA=40$\degr$.  It shows a
maximum velocity amplitude of $\pm$ 70 km s$^{-1}$, which is
significantly lower than that for H90b and H90d.  The mean PA of the
major axis given by the gas kinematics, 40$\degr$ $\pm$ 5$\degr$,
differs by 60$\degr$ from the corresponding stellar kinematic major
axis of 100$\degr$ (as given by L94; see also Section 4.1).

   The NE side of the velocity curve of H90c is not as flat as the SW
side. This may be due to a local starburst in the NE of the galaxy, 
which induces proper motions (see Fig. 2).

  The center of the galaxy as measured from the continuum image
coincides with the monochromatic center within the seeing disk ($\sim$
1$\arcsec$). The kinematic center, however, is situated 2$\arcsec$ to
the west of the other two centers.  The gas systemic velocity, as
measured from the VF, is 2785 $\pm$ 15 km s$^{-1}$.  Hickson (1989) and
de Carvalho et al. (1997) found systemic velocities of 2696 $\pm$ 24 km
s$^{-1}$ and 2497 $\pm$ 12 km s$^{-1}$ respectively, from measurements
using absorption lines.  The central velocity dispersion of the galaxy,
as measured by the full width of half maximum (FWHM) of the gas profile
is 80 km s$^{-1}$ $\pm$ 15 (after deconvolution by the instrumental
profile of FWHM $\sim$ 35 km s$^{-1}$).  This central value decreases
to 63 km s$^{-1}$ $\pm$ 16 towards the external regions.

The integrated flux calculated for the total gas extension is
F(H$\alpha$)=1.45 $\pm$ 0.12 $\times$ 10$^{-14}$ erg s$^{-1}$
cm$^{-2}$.  Using the formula given by Osterbrock (1974) with an
electronic density of 1000 cm$^{-3}$, we have derived a mass of ionized
gas of M$_{HII}$=0.45 $\pm$ 0.04 $\times$ 10$^4$  M$_{\odot}$.  This
may, however, be an upper limit on the gas mass, since the value we
take for the electronic density is also an upper limit. 

We used the formula from Lequeux (1983) (with f=1) to determine the
total mass of the galaxy.  We have obtained M$_{tot}$= 0.48 $\times$
10$^{10}$ $\pm$ 0.13 M$_{\odot}$ (see Table 2).

\placefigure{fig1}

\placefigure{fig2}

\placefigure{fig3}

\placefigure{fig4}

\subsection{H90b  (NGC 7176)}

H90b has been classified as an E0 galaxy by Hickson (1993) and as an E+pec
galaxy in the RC3 catalogue (De Vaucouleurs et al. 1991). As for HCG 90C, the
round shape of this object is a problem to determine the PA of the
major axis.  Mendes de Oliveira (1992) reported a value of
55$\degr$ $\pm$ 5$\degr$ from the analysis of an R image of the galaxy
which is consistent with the values of 53$\degr$ $\pm$
10$\degr$ (see Fig. 1) and 55$\degr$ $\pm$ 10 determined from
the continuum and monochromatic images respectively.

In Figs. 4a and 4b we show the monochromatic and velocity maps for the
ionized gas component of H90b.  The warm gas in H90b can be measured
out to a radius of $\sim$ 20$\arcsec$ which represents 1.3 $r_{eff}$ or
3.2 h$^{-1}$ kpc (using $r_{eff}$ given by L94).  The iso-velocities of
the VF are fairly regular in the east side of the galaxy. In the SW
direction, however, the situation is much more confusing.  The
distribution of the ionized gas is very clumpy.  For this reason it is
difficult to determine its center.  An excess of gas is observed in the
SW part of the galaxy, towards H90d.

 Fig. 5 shows the variation of the PA of the major axis with radius,
derived from the VF. The kinematic PA of the major axis raises
continuously from $\sim$ 0$\degr$ to 100$\degr$ at a 17$\arcsec$-radius.
The mean PA of the major axis given by the gas kinematics,
60$\degr$ $\pm$ 5$\degr$, differs by 70$\degr$ from the corresponding
stellar kinematic major axis of $\sim$ 130$\degr$ (as given by L94; see
also Section 4.1).

  We find that the FWHM of the emission-line profiles of H90b are
slightly lower than those for H90c. We find a mean FWHM of $\sim$ 60
$\pm$ 10 km s$^{-1}$ in the center (after correction for an
instrumental profile of FWHM $\sim$ 35 km s${-1}$).  This value
decreases with distance to the center to 47 $\pm$ 10 km s$^{-1}$
at a 15$\arcsec$-radius.

  Fig. 6 presents the VC derived from the velocity field of H90b with a
PA=60$\degr$, our best determination of the kinematic major axis of the
galaxy. This curve was built using the kinematic center (situated
3.3$\arcsec$ to the NE of the continuum center) and its corresponding
velocity of 2540 $\pm$ 12 km s$^{-1}$. As a comparison, Hickson et al.
(1992) and de Carvalho et al. (1997) give systemic velocities for H90b
of 2525 $\pm$ 29 km s$^{-1}$ and 2511 $\pm$ 14 km s$^{-1}$
respectively,  measured from absorption lines.  H90b has an
asymmetric VC, rare among isolated galaxies.  The maximum velocity
observed is +75 km s$^{-1}$ and the minimum is --110 km s$^{-1}$.

   The total integrated H$\alpha$ flux inside a 18$\arcsec$-radius is
7.8 $\pm$ 0.15 $\times$ 10$^{-15}$ erg s$^{-1}$ cm$^{-2}$, which gives
a gas mass of M$_{HII}$=0.24 $\pm$ 0.04 $\times$ 10$^4$ M$_{\odot}$
(assuming an electronic density of 1000 cm$^{-3}$).  This gives an
upper limit for the warm gas mass.

   The total mass of the galaxy,  obtained using the formula of Lequeux
(1983), was found to be M$_{tot}$=0.82 $\times$ 10$^{10}$ $\pm$ 0.25
M$_{\odot}$ (Table 2).

\subsection{H90d (NGC 7174)}

\placefigure{fig5}

This object has been classified as an Irregular galaxy by Hickson
(1989) and as an early-type spiral by the RC3 catalogue (de Vaucouleurs
et al. 1991). The PA of its major axis, as derived from the continuum
image is 85$\degr$ $\pm$ 10$\degr$ and 75 $\degr$ $\pm$ 5$\degr$
from the monochromatic image.

   Figs. 7a and 7b present respectively the monochromatic image and the
velocity field.  Emission is measured out to a radius of $\sim$
15$\arcsec$. Fig. 7a shows a gas extension in the east side of the
galaxy, towards H90b.  Fig. 7b shows disturbed iso-velocities in the NE
and SW direction.

   The monochromatic center coincides with the center of the continuum
image (within the seeing disk). Both are, however, separated
from the kinematic center by approximately 3$\arcsec$ (Fig. 7b).
Our best determination of the mean PA of the major-axis is 75$\degr$ $\pm$
5$\degr$ (as measured from the overall geometry of the kinematic map).

The PA of the monochromatic image does not change significantly with
radius while the PA of the kinematic major axis does (see Fig. 5).  The
variation of the PA of the kinematic major axis for H90d shown in Fig.
5 suggests the presence of a strong warp.

  The systemic velocity is determined from the  VF to be 2635
$\pm$ 15 km s$^{-1}$, similar to the systemic velocity given by de
Carvalho et al.  (1997) of v$_{sys}$=2659 $\pm$ 9 km s$^{-1}$, also
determined from emission lines. 

  The VC shown in Fig. 6 was derived from the VF, along the kinematic
major axis, PA=75$\degr$.  It reaches a maximum velocity of +200 km
s$^{-1}$ at 15$\arcsec$ from the nucleus, in the SW side.  In the NE
side it reaches a minimum of --150 km s$^{-1}$ at a
28$\arcsec$--radius. The motion is not axisymmetric. On the SW side of
the curve, the VC is perturbed as displayed by two of the
iso-velocities present in that region. A possible explanation is the
presence of a dust lane which affects the VF.

  The velocity dispersion at the position of the kinematic and
continuum centers are respectively 110 $\pm$ 10 km s$^{-1}$  and
120 $\pm$ 10 km s$^{-1}$ (after correction for an instrumental profile
of FWHM $\sim$ 35 km s$^{-1}$.  Further to the east it decreases to 90
$\pm$ km s$^{-1}$.

  The total gas flux derived within a 16$\arcsec$-radius is
F(H$\alpha$)=2.39 $\pm$ 0.12 $\times$ 10$^{-14}$ erg s$^{-1}$
cm$^{-2}$, which gives an upper limit on the gas mass of M$_{HII}$=1.0
$\pm$ 0.04$\times$ 10$^4$ M$_{\odot}$.

   We obtain a total mass for this galaxy of M$_{tot}$ = 2.4 $\times$
10$^{10}$ $\pm$ 0.2 M$_{\odot}$ (see Table 2).

\placefigure{fig6}

\placefigure{fig7}

\placefigure{fig8}

\section{General discussion of the results}

\subsection{Comparison with the Stellar Kinematics of the Galaxies}

L94 have carried out a study of the stellar kinematics of the HCG 90
galaxy members.  They concluded that H90c and H90d are in interaction
and H90b and H90d are not. This last point is contrary to our results.

  We give details of the comparison between the stellar and gaseous 
kinematics for each of the galaxies below.

\begin{itemize} 

\item H90c: 

  L94 presented stellar velocity curves for H90c along slits placed at
PA=130$\degr$ and PA=147$\degr$.  We present here a comparison at only
one of these position angles, since the kinematics of the gas does not
change significantly between these two positions.

 Fig. 3b presents (with crosses) the VC derived from the H$\alpha$
kinematics, along a PA=130$\degr$, which coincides with the gas kinematic
minor axis of the galaxy.  We would expect to find a flat curve, if no
motion was present along the minor axis. We detect, however, a small
velocity gradient due to:  1) two inconsistent isovelocities in the SE
part of the VF (see Fig. 2b) which produces positive velocities on the
VC (+50 km s$^{-1}$ at 13$\arcsec$) and 2) on the NW side, the cross
section at PA=130$\degr$ includes two twisted iso-velocities which
produces negative velocities on the VC (--50 km s$^{-1}$ at 18$\arcsec$).
These features are produced by 
iso-velocities with S/N $\sim$
3 and 8 respectively. 

 We have overplotted in Fig. 3b the stellar VC (from L94, his Fig. 3a)
determined at the same PA of 130$\degr$.  The
shapes of the gas and stellar curves are very different.  For the gas
kinematics we do not see the ``U shape''  described by the stellar
kinematics.  Another striking difference is the lack of a flat portion
in the velocity curve along the major axis of the stellar component and
the presence of a irregular
velocity gradient that goes up to 300 km s$^{-1}$
with no regular motion, contrary to what is seen at a cut along the
kinematic major axis of the gas component (shown in Fig. 3a). According
to L94 the stellar major axis of the galaxy may be at 100$\degr$
(corresponding to a rotating axis of 10$\degr$). Therefore, the cut at
PA=130$\degr$ is close to the major axis defined by the stellar
kinematics and should show a large velocity gradient, as it indeed
does.

  The stellar and gaseous motions are
strongly decoupled (by about 100$\degr$--40$\degr$=60$\degr$). 
The kinematics of the gaseous component is more well
behaved than that of the stellar component.  This is evidence that
this early-type galaxy contains a gas disk with well ordered
circular motion. 

\item H90b+d: 

  These two galaxies have strongly overlapping isophotes. Their continuum
centers are separated by only 25$\arcsec$ (4 h$^{-1}$ kpc).

  The non-interaction between H90b and H90d claimed by L94 was based on
the following evidence:  1) no disturbances were found in the internal
stellar velocity curves of the two galaxies along an axis which joins
their centers (PA=68$\degr$) and 2) the parabolic velocity calculated for
these two galaxies is twice their systemic velocity difference.
Point (1) is discussed below and point (2) is
discussed in the next section.

The stellar velocity curve obtained
along the line joining the centers of H90b and H90d (PA=68$\degr$)
was given by L94 (their Fig. 7a) and it can be directly compared to our
Fig. 8.  The gas velocity curve shown in Fig. 8 was obtained by doing a
cross-section of the gas velocity maps of H90b and H90d at PA=68$\degr$
and using the continuum center and the systemic velocity of H90b as a
reference point (position 0,0 in Fig. 8), in order to be consistent
with L94. Since it was not specified in L94's paper which systemic
velocity was used to draw their Fig. 7a, we will make a comparison
of the shape and extension of the curves only, not of the
absolute values for the maximum velocities.

   The first obvious difference when comparing the gas and stellar
curves is the contrast between the {\it continuous stellar} rotation
curves of H90b and H90d with the {\it break} in velocity space between the two
corresponding {\it gaseous} components.  This may occur, at least in part,
because our observations are made at higher resolution than L94's
observations. 

  We also see a strong discrepancy between the shape of the gaseous and
stellar components for H90b (around position 0,0 in Fig. 8). The stellar
component is almost flat while the gas curve displays a large velocity
gradient. This is easily understood if we remember that the position
angle PA=68$\degr$, along which the measurements are plotted in Fig. 8,
is very close to the kinematic major axis of the gas component of H90b
(PA=60$\degr$, see Table 2) but almost perpendicular to the kinematic
major axis of the stellar component (PA = 130$\degr$, see L94).
Therefore we would expect the velocity curves for the two components of
H90b to have very different shapes, as is indeed observed.

 The shapes of the stellar and gaseous velocity curves of H90d, in
contrast, are in general agreement (in the region where a comparison is
possible, i.e., to the SW side of the galaxy).  This was expected given
that the major axes of the gas and stellar components of H90d have the
same orientation (PA$\sim$75$\degr$).

  Further comparison of the gaseous and stellar curves show that the
stellar measurements do not go as far to the NE as the gas velocity
curve does.  The opposite is true for the SW region. The more extended
stellar velocity curve on this side may be due to a light extension
detected by L94 but not seen in the ionized gas (perhaps because of too
low S/N of our data in that region).

   A plot of our data for H90b along a PA=130$\degr$ (not shown here)
confirms that this is indeed the minor axis of the gas kinematic disk
(the curve is flat). This contrasts with the curve drawn at the same PA
(in Fig. 3a of L94) for the stellar component, which displays a
velocity gradient.  It is clear from the comparison described above
that the kinematics of the stellar and gaseous components of H90b
are strongly decoupled (by about
130$\degr$--60$\degr$=70$\degr$).

  In order to convince the reader that the gaseous velocity profiles of 
H90b+d are separated and can be measured in the region of overlap we
plot in Fig.  9 part of the 2-D H$\alpha$ map.  The upper
pannel indicates with a bold square the region within which the
H$\alpha$ profiles are shown in the lower pannels.  The lower left and
right pannels show respectively the measured profiles and the best
gaussian fits to the data (of FWHM = 63 km s$^{-1}$).  Each small
square in the lower pannels correspond to a 0.91$\arcsec$ $\times$
0.91$\arcsec$ -- pixel and a velocity interval of 380 km s$^{-1}$.  We
can clearly measure both components of H90b and H90d in the overlaping
region.

\placefigure{fig9}

\item Parabolic velocities 

L94 argued that the system H90b$+$d is not in interaction.  One of
the reasons presented was the fact that the parabolic velocity is much
larger than the difference between the two systemic velocities.  We did
the same calculation using the gas parameters (the gas central velocity
and the total mass determined from the gas kinematics and the
virial theorem) and we find as
parabolic velocities  V$_{para}$=130 $\pm$ 45 km s$^{-1}$ and
 V$_{para}$=230 $\pm$ 55 km s$^{-1}$ for the systems H90c+d and H90b+d 
 respectively. This
can be compared with the central gas systemic velocity differences of
150 $\pm$ 30 kms $^{-1}$ and 95 $\pm$ 30 km s$^{-1}$ respectively. 
Errors bars are  large,
 and therefore the discrepancy between the parabolic velocities and velocity
 differences may be not significant.

\end{itemize}

\subsection{Misclassification of two members}

  H90b and H90c have been previously classified as elliptical
galaxies.  Our kinematic study of the gas component,
however, show that these two objects may be S0 galaxies.

%We show in Figs. 10a and 10b the surface-brightness profiles (from
%the H$\alpha$ images) of H90b, c and d plotted as a function of radius
%and as a function of radius to the power 0.25. These plots show that
%the H$\alpha$ intensity distributions for the three galaxies are not
%good fits to either exponential or r$^{1/4}$-law profiles.  H90b and
%H90d have smooth profiles while H90c has a more irregular distribution. H90c profile
%shows a little bump at 6.6" from the center, this corresponds to the small
%region located to the north at the same radius and showed in Fig.2a.
%distribution. For H90b there is an excess of gas at 15.5$\arcsec$ that
%can be seen in the profile of Figs. 10a and 10b and also in the
%monochromatic map of Fig. 4a.

%\placefigure{fig10}

   Combining the gas and stellar kinematic data,
we conclude that H90b and H90c may be S0 galaxies.  
According to L94, the ratio V/$\sigma$ of H90c, as measured from the
stellar component, is 0.7 for an apparent ellipticity of 0.26. This
places H90c in the V/$\sigma$
{\it vs.} ellipticity diagram of Davies et al. (1987) in the domain of
the rotationally supported objects.  Our gas study leads to the same
conclusion, with  V/$\sigma$ =0.87 for an apparent ellipticity  of
0.1.  Likewise, for H90b, the stellar V/$\sigma$ is 0.6, for an
apparent ellipticity of 0.28, and for the gas component the V/$\sigma$ is
1.6  for an apparent ellipticity of 0.6.  Therefore, both H90b and H90c
may be rotationally supported objects.

  We conclude from the position of H90b and H90c in the V/$\sigma$ {\it vs.}
ellipticity diagram   that
these are misclassified S0 galaxies.

\subsection{Ionized gas mass to K-band luminosity ratio}

Fig. 10 presents the ratio of mass of HII gas to K band luminosity as a
function of radius from the center of the galaxy for galaxies H90b, c
and d.  The HII gas masses have been derived from the calibrated
monochromatic images and transformed into mass densities in M$_{\odot}$
pc$^{-2}$. The K band luminosity profiles come from unpublished
photometry by Mendes de Oliveira, transformed into surface density in
L$_{\odot}$ pc$^{-2}$.  The absolute scale for the K photometry is not
accurate but the relative differences in the shapes and intensities of
the curves in Fig. 11 are not affected by this uncertainty given that
the zero point of the K photometry is identical for the three
galaxies.  Tables 3, 4 and 5 give details of the plots shown in Fig.
10.

  The two early-type galaxies, H90b and H90c, present a lower mean
M$_{HII}$/L$_{K}$ ratio than H90d. The M$_{HII}$/L$_{K}$ of H90c
and H90d raise regularly with approximately the same slope.  The slope
of the curve of H90b is, however, very different from those for the two
other galaxies. For H90b the M$_{HII}$/L$_{K}$ increases rapidly from
--7.0 M$_{\odot}$ / L$_{\odot}$ at a radius of 4$\arcsec$ to --5.75
M$_{\odot}$ / L$_{\odot}$ at 15$\arcsec$ from the center. After that it
decreases slowly.  The rapid increase corresponds to the gas excess
present in the monochromatic map of H90b (see Fig. 4a). Between radii
15$\arcsec$ and 20$\arcsec$, the M$_{HII}$/L$_{K}$ ratio increases
for H90d while it decreases for H90b. An analysis made by Hibbard et al.
(1995) of the HI content of several merger remnants showed that an
increase of the M$_{HI}$/L$_{K}$ in the outer parts of an
interacting system is a strong signature of a late-stage encounter.  We
cannot be as affirmative here, mainly due to the extension of the gas
which is very small compared to that for the HI gas.  What we can say,
however, is that we detect a significant difference in the behaviour of
the M$_{HII}$/L$_{K}$ {\it vs.} radius curve for H90b, as compared to
those for H90c and H90d.

In a scenario where H90d is considered as a gas reservoir and H90b is
the ``recipient'', it seems plausible to find a higher amount of gas
for H90d and a higher M$_{HII}$/L$_{K}$ ratio in the extreme part of
H90b, where the gas exchange is taking place.

\placefigure{fig10}

\placetable{tbl-3}

\placetable{tbl-4}

\placetable{tbl-5}

\subsection{On-going interaction signatures}

\begin{itemize}

\item Insights from the study of the kinematics and gas content

The values listed in Table 2 show that the total ionized gas mass in
H90d is four times higher than that for H90b and twice as high as
that of H90c. The values found for the two early-type galaxies,
however, even if small, are consistent with previous studies of the
interstellar medium (ISM) of elliptical and lenticular galaxies
in other environments (Goudfrooij et al. 1994, Macchetto et al. 1996).
What is anomalous in the early-type galaxies H90b and H90c is not their
gas content but their gas kinematics, as summarized below.

- H90c: 

The strong decoupling between the warm gas and the stellar major axes
for this galaxy and the contrast between the well behaved kinematics
of the gas and the `U-shape' kinematics of the stars along its
kinematic major axes suggest that the
gas disk of H90c is of external origin. It may be the result of an
interaction. The gas is relaxed while the stars still feel the effects
of a recent interaction.

- H90b+d:

The VFs and the monochromatic maps of this system strongly suggest that
there is an interaction in progress taking place between these two
galaxies.  The VFs of the two galaxies are twisted and the VCs show
non-axisymmetric motions.  The monochromatic maps show evidence that
there is a gas bridge between these two galaxies:  there is an excess of
ionized gas in the SW side of H90b, towards H90d, and the gas
distribution is non-axisymmetric with an extension to the NE side of
H90d, in the direction of H90b.

  L94 concluded that there is no strong observational evidence for an
interaction in progress between H90b and d from the stellar
kinematics.  This difference between the kinematics of the gas and of
the stars can arise, for an example, if the gas component responds
more quickly to gravitational perturbations than the stellar component. In
this case the gas kinematics may show evidences of interaction that do
not have counterparts in the stellar kinematics. 

  The hypothesis that H90b+d form an interacting system is enforced by
the observation that in binary galaxies the interaction is more
efficient when the encounter is prograde (Alladin
\& Narasimhan 1982), which is the case for this system.

\item Insights from other ISM phases

Other components of the ISM of the galaxy members  can help 
determining the evolutionary stage of the compact group HCG 90.

  Several studies show that HCG 90 has diffuse X-ray emission (Ponman et
al.  1996, Mulchaey \& Zabludoff 1998). The emission is not centered on
any group member, suggesting that HCG 90 may be a young, dynamically
evolving group.

  H90a is not included in this study but HI observations for this galaxy
are available by Oosterloo and Iovino (1997).  Located at 8$\arcmin$ or
77 h$^{-1}$ kpc to the north of the other three members, the velocity
difference between H90a and the rest of the group is only 100 km
s$^{-1}$.  The morphology of the HI emission is very disturbed, showing
an extension in the direction of the remaining group members (mainly to
HCG 90C).  Oosterloo and Iovino (1997) concluded that the HI morphology
for this galaxy suggests a strong tidal interaction with the other
group members.

  CO emission has been detected in HCG 90  by Verdes-Montenegro et al.
(1998) mainly centered on H90d. However Huchtmeier \& Tammann (1992)
show, from smaller beam data, that H90b also displays significant CO
emission. The case of H90c is still not resolved. Verdes-Montenegro et
al. (1998) indicate that this member presents only marginal CO emission,
but they suggest that observations with a smaller beam may be able
to confirm their detection.

\end{itemize}

\subsection{Possible evolutionary scenario}

  We show in this paper several pieces of evidence that HCG 90 is a
dynamically evolving group. The three galaxies are in
interaction and will possibly merge into one single object.

  We suggest the following scenario: H90d is the warm gas reservoir of
the group. H90c and d have experimented a past interaction with gas
exchange. The gas acquired by H90c has already settled and relaxed
(as suggested by the well behaved shape of the gas velocity curve of
H90c) but the effects of the interaction can still be visible in the
morphology of the two galaxies and their stellar kinematics (L94).  On
the other hand, H90d is now in the process of fueling H90b with gas,
a process that will possibly result in a major
merger.

  Our observations provide important constraints for the modelling of
the kinematics of merging group galaxies.

\acknowledgements

The authors thank Chantal Balkowski for initializing the project and
Jean-Luc Gach for mounting the instrument on the telescope and helping
during the observations.  P. Amram and J. Boulesteix thank the
department of Astronomy at the IAG in S\~ao Paulo for the hospitality
during their visit, in connection with this project.  H.  Plana
acknowledges the financial support of the Brazilian FAPESP, under
contract 96/06722-0.  C. Mendes de Oliveira acknowledges the financial
support from the Alexander von Humboldt foundation.

\clearpage

%
%Table 1 Journal of obs
%

\begin{deluxetable}{lrr}
\tablenum{1}
\tablewidth{0pc}
\tablecaption{Journal of Perot-Fabry observations}
\tablehead{
\colhead{} & \colhead{Compact Group of Galaxies Hickson 90 } & \colhead{}}
\startdata

Observations & Telescope & ESO 3.6m \nl 
	     & Equipment & CIGALE @ Cassegrain focus \nl 
	     & Date & August, 28th 1995 \nl 
	     & Seeing & $\sim$ 1.2" \nl 
Interference Filter &   Central Wavelength & 6624 \AA \tablenotemark{1}  \nl 
		    &   FWHM & 10 \AA  \tablenotemark{2} \nl 
		    &   Transmission & 0.6 \tablenotemark{1}  \nl 
Calibration & Neon Comparison light & $\lambda$ 6598.95 \AA \nl 
Perot--Fabry & Interference Order & 796 @ 6562.78 \AA \nl 
		 & Free Spectral Range at H$\alpha$ & 380 km s$^{-1}$ \nl 
		 & Finesse at H$\alpha$ & 12 \nl 
                 & Spectral resolution at H$\alpha$ & 18750 at the sample step \nl
Sampling & Number of Scanning Steps & 24 \nl 
	 & Sampling Step & 0.35 \AA\ (16 km s$^{-1}$) \nl 
	 & Total Field & 230''$\times $230'' (256$\times $256 px$^2$) \nl 
	 & Pixel Size & 0.91''  \nl 
Exposures times & Total exposure & 2 hours \nl 
		& Elementary scanning exposure time & 5 s per channel \nl 
		& Total exposure time per channel & 300 s\nl

\tablenotetext{1} {For a mean beam of 2.7$^o$ }
\tablenotetext{2} {For a mean beam inclination of 2.7$^o$ }

\enddata

\end{deluxetable}

\clearpage

%
%Table 2 Physical parameters
%

\begin{deluxetable}{lccc}
\tablenum{2}
\tablewidth{0pc}
\tablecaption{Physical Parameters of the Galaxies}
\tablehead{
\colhead{Name} & \colhead{H90b} & \colhead{H90c}  & \colhead{H90d} }
\startdata
Other names  & NGC 7176 & NGC 7173 & NGC 7174\nl

$\alpha$ (1950) $^1$ & 21$^{h}$59$^{m}$14.1$^{s}$ &   
21$^{h}$59$^{m}$08.8$^{s}$&  21$^{h}$59$^{m}$11.6$^{s}$\nl

$\delta$ (1950) $^1$ & --32\arcdeg11'36.4" & --32\arcdeg11'30.4" & 
--32\arcdeg11'26.3"\nl

Morphological type (Hickson/RC3) $^1$ & E0/E+pec & E0/E+pec & 
Im/Sab\nl

B$_{T_C}$ $^1$ & 12.57 &12.73 & 12.81\nl

Systemic heliocentric velocity/$^1$ (km s$^{-1}$)  & 2525 $\pm$ 29 
&  2696 $\pm$ 24 & 2778 $\pm$ 29\nl

Gas central velocity/$^7$ (km s$^{-1}$) &   2540 $\pm$ 12  &  2785 $\pm$ 15 
&  2635 $\pm$ 15   \nl

D(Mpc)$^2$ & 33.15 & 33.15 & 33.15 \nl

F(H$\alpha$)10$^{-14}$ erg s$ ^{-1}$ cm$^{-2}$/$^3$ & 0.78 $\pm$ 0.15 & 1.45 
$\pm$ 0.12 & 2.39 $\pm$ 0.12 \nl

Log L(H$\alpha$) erg s$ ^{-1}$ & 37.85 $\pm$ 0.20 & 38.2 $\pm$ 0.02 & 
38.41 $\pm$ 0.05 \nl

M$_{HII}$ ($\times$ 10$^4$ M$_{\odot}$) & 0.24$\pm$ 0.04 & 0.45 $\pm$ 0.04 & 
1.0 $\pm$ 0.04\nl

Vrot$_{max}$ (km s$^{-1}$)/$^4$ & 122 $\pm$ 12 & 85 $\pm$ 12 & 220 $\pm$ 18\nl

R$_{max}$ (kpc)/$^5$ & 2.40 $\pm$ 0.25 & 2.90 $\pm$ 0.25 & 2.20 $\pm$ 0.25\nl

M$_{tot}$ ($\times$ 10$^{10}$ M$_{\odot}$)/$^6$ & 0.82 $\pm$ 0.25 & 0.48 $\pm$ 0.13 & 
2.4 $\pm$ 0.2\nl

Major axis (gas kinematics)/$^7$ & 60\arcdeg $\pm$ 10 & 40\arcdeg $\pm$ 5 & 75\arcdeg 
$\pm$ 10\nl

Major axis (stellar kinematics)/$^8$  & 130\arcdeg & 100\arcdeg
 & 68\arcdeg/74\arcdeg \nl

Major axis (H$\alpha$ image)/$^3$ & 55\arcdeg $\pm$ 10 & 51\arcdeg $\pm$ 20 & 
85\arcdeg $\pm$ 5\nl

Inclination of the gaseous disk $^7$ & 54\arcdeg $\pm$ 3 & 
55\arcdeg $\pm$ 3 & 55\arcdeg $\pm$ 5\nl

FWHM of central profiles (km s$^{-1}$) & 60 $\pm$ 16 & 80
$\pm$ 15 & 110 $\pm$ 8 \nl

\tablerefs{
$^1$ Hickson (1993);
$^2$ from Faber et al. (1989); 
$^3$ from calibrated monochromatic maps;
$^4$ maximum rotation velocities (corrected for the gas-disk inclination);
$^5$ radius at maximum velocity, using distance for the group 
from Faber et al. 1989;
$^6$ total mass of the galaxy, determined using Lequeux's (1983) formulae (with
f=1.);
$^7$ derived from the gas velocity map;
$^8$ from L94
}
\enddata
\end{deluxetable}

\clearpage

%
%Table 3 
%

\begin{deluxetable}{c c c c c}
\tablenum{3}
\tablewidth{0pc}
\tablecaption{H90b surface brightness and surface density. }
\tablehead{
\colhead{Radius$^1$} & \colhead{K$^2$} &\colhead{Log[$\sigma(K )$]$^3$} & \colhead{Log[$\sigma(HII)$]$^4$} &
\colhead{Log[$\sigma(HII) \over \sigma(K )$]$^5$} \nl
\colhead{\arcsec} & \colhead{mag/"$^2$} & \colhead{L$_{\odot}.pc^{-2}$} & 
\colhead{M$_{\odot}.pc^{-2}$} & \colhead{M$_{\odot}$/L$_{\odot}$}} 
\startdata
     4.19  &     17.77   &    3.626   &   -3.310  &    -6.936    \nl 
     4.82  &     18.00   &    3.533   &  -3.319   &   -6.852     \nl
     5.37  &     18.23   &    3.442   &   -3.384  &    -6.826    \nl
     7.83  &     18.94   &    3.157   &   -3.394  &    -6.551    \nl
    15.02  &     20.34   &    2.600   &   -3.102  &    -5.702    \nl
    15.47  &     20.43   &    2.562   &  -3.102   &   -5.664    \nl
    16.38  &     20.53   &   2.521    &  -3.132   &   -5.653    \nl
    16.47  &     20.53   &    2.521   &   -3.136  &    -5.657   \nl
    16.74  &     20.59   &    2.500   &   -3.201  &    -5.701   \nl
    17.29  &     20.65   &    2.475   &   -3.250  &    -5.725   \nl
    17.47  &     20.70   &    2.454   &   -3.272  &    -5.726   \nl
    18.20  &     20.78   &   2.422    &  -3.396   &  -5.818    \nl
    18.93  &     20.90   &    2.376   &  -3.425   &   -5.801   \nl
    19.47  &     20.99   &    2.340   &   -3.486  &    -5.826  \nl
    20.38  &     21.11   &    2.289   &   -3.543  &    -5.832  \nl
\tablerefs{
$^1$ Radius from the center in arcsec; \\
$^2$ Surface brightness in Mag/"$^2$; \\
$^3$ Logarithm of light surface density in L$_{\odot}$/pc$^2$; \\
$^4$ Logarithm of gas surface density in M$_{\odot}$/pc$^2$; \\
$^5$ Logarithm of gas mass to K bandluminosity ratio in M$_{\odot}$/L$_{\odot}$.
}

\enddata

\end{deluxetable}

\clearpage

%
%Table 4 
%

\begin{deluxetable}{c c c c c}
\tablenum{4}
\tablewidth{0pc}
\tablecaption{H90c surface brightness and surface density}
\tablehead{
\colhead{Radius$^1$} & \colhead{K $^2$} &\colhead{Log[$\sigma(K )$]$^3$} & \colhead{Log[$\sigma(HII)$]$^4$} &
\colhead{Log[$\sigma(HII) \over \sigma(K )$]$^5$} \nl
\colhead{\arcsec} & \colhead{mag/"$^2$} & \colhead{L$_{\odot}.pc^{-2}$} & 
\colhead{M$_{\odot}.pc^{-2}$} & \colhead{M$_{\odot}$/L$_{\odot}$}} 
\startdata
	
       2.57   &   17.15       &      3.825   & -3.004   &-6.829  \nl
       2.81   &    17.36      &       3.792   &-3.024   &-6.816  \nl
       3.21   &    17.55      &       3.716   &-3.054   & -6.770  \nl
       3.37   &    17.63      &       3.681   &-3.067   &-6.748  \nl
       3.60   &    17.75      &       3.636   &-3.085   &-6.721  \nl
       3.81   &    17.83      &       3.602   &-3.103   &-6.705  \nl
       4.11   &    17.91      &       3.571   &-3.129   &  -6.700  \nl
       4.54   &    18.08      &       3.504   &-3.165   &-6.669  \nl
       5.24   &    18.39      &       3.376   &-3.220    &-6.596  \nl
       5.81   &    18.55      &      3.314    &-3.261   & -6.575  \nl
       6.45   &    18.77      &      3.225    &-3.307   & -6.532  \nl
       7.67   &    19.09      &      3.097    &-3.388   &-6.485  \nl
       10.73   &    19.86      &       2.792   &-3.545   & -6.337  \nl
       15.25   &    20.79      &      2.417    & -3.711  &-6.128 \nl
       19.98   &    21.69      &        2.056  &  -3.86   &-5.916  \nl
\tablerefs{
$^1$ Radius from the center in arcsec; \nl
$^2$ Surface brightness in Mag/"$^2$; \nl
$^3$ Logarithm of light surface density in L$_{\odot}$/pc$^2$; \nl
$^4$ Logarithm of gas surface density in M$_{\odot}$/pc$^2$; \nl
$^5$ Logarithm of gas mass to K band luminosity ratio in M$_{\odot}$/L$_{\odot}$.
}
\enddata

\end{deluxetable}

\clearpage

%
%Table 5 
%

\begin{deluxetable}{c c c c c}
\tablenum{5}
\tablewidth{0pc}
\tablecaption{H90d surface brightness and surface density}
\tablehead{
\colhead{Radius$^1$} & \colhead{K $^2$} &\colhead{Log[$\sigma(K )$]$^3$} & \colhead{Log[$\sigma(HII)$]$^4$} &
\colhead{Log[$\sigma(HII) \over \sigma(K )$]$^5$} \nl
\colhead{\arcsec} & \colhead{mag/"$^2$} & \colhead{L$_{\odot}.pc^{-2}$} & 
\colhead{M$_{\odot}.pc^{-2}$} & \colhead{M$_{\odot}$/L$_{\odot}$}} 
\startdata
       1.82    &   17.64   &     3.680  &    -2.517   &   -6.192  \nl
       2.00    &   17.69   &     3.660  &    -2.522   &   -6.182  \nl
       2.18    &   17.74   &     3.640  &    -2.523   &   -6.163  \nl
       2.37    &   17.75   &     3.635  &    -2.553   &   -6.188  \nl
       2.55    &   17.84   &     3.600  &    -2.563   &   -6.163  \nl
       2.82    &   17.78   &     3.624  &    -2.579   &   -6.203  \nl
       3.09    &   17.82   &     3.607  &    -2.603   &   -6.210  \nl
       3.37    &   17.90   &     3.576  &    -2.627   &   -6.203  \nl
       9.65    &   18.94   &     3.159  &    -2.674   &   -5.833  \nl
      14.38    &   19.47   &     2.947  &    -2.728   &   -5.675  \nl
      15.11    &   19.55   &     2.914  &    -2.776   &   -5.690  \nl
      15.83    &   19.69   &     2.860  &    -2.805   &   -5.665  \nl
      19.66    &   20.66   &     2.472  &    -2.866   &   -5.338  \nl
      20.84    &   20.82   &     2.407  &    -2.923   &   -5.330  \nl
      21.29    &   20.89   &     2.380  &    -2.996   &   -5.376  \nl
\tablerefs{
$^1$ Radius from the center in arcsec; \nl
$^2$ Surface brightness in Mag/"$^2$; \nl
$^3$ Logarithm of light surface density in L$_{\odot}$/pc$^2$; \nl
$^4$ Logarithm of gas surface density in M$_{\odot}$/pc$^2$; \nl
$^5$ Logarithm of gas mass to K band luminosity ratio in M$_{\odot}$/L$_{\odot}$.
}

\enddata

\end{deluxetable}

\clearpage

\begin{table}
\dummytable\label{tbl-1}
\end{table}

\begin{table}
\dummytable\label{tbl-2}
\end{table}

\begin{table}
\dummytable\label{tbl-3}
\end{table}

\begin{table}
\dummytable\label{tbl-4}
\end{table}

\begin{table}
\dummytable\label{tbl-5}
\end{table}

%
%Fig. 1
%

\figcaption[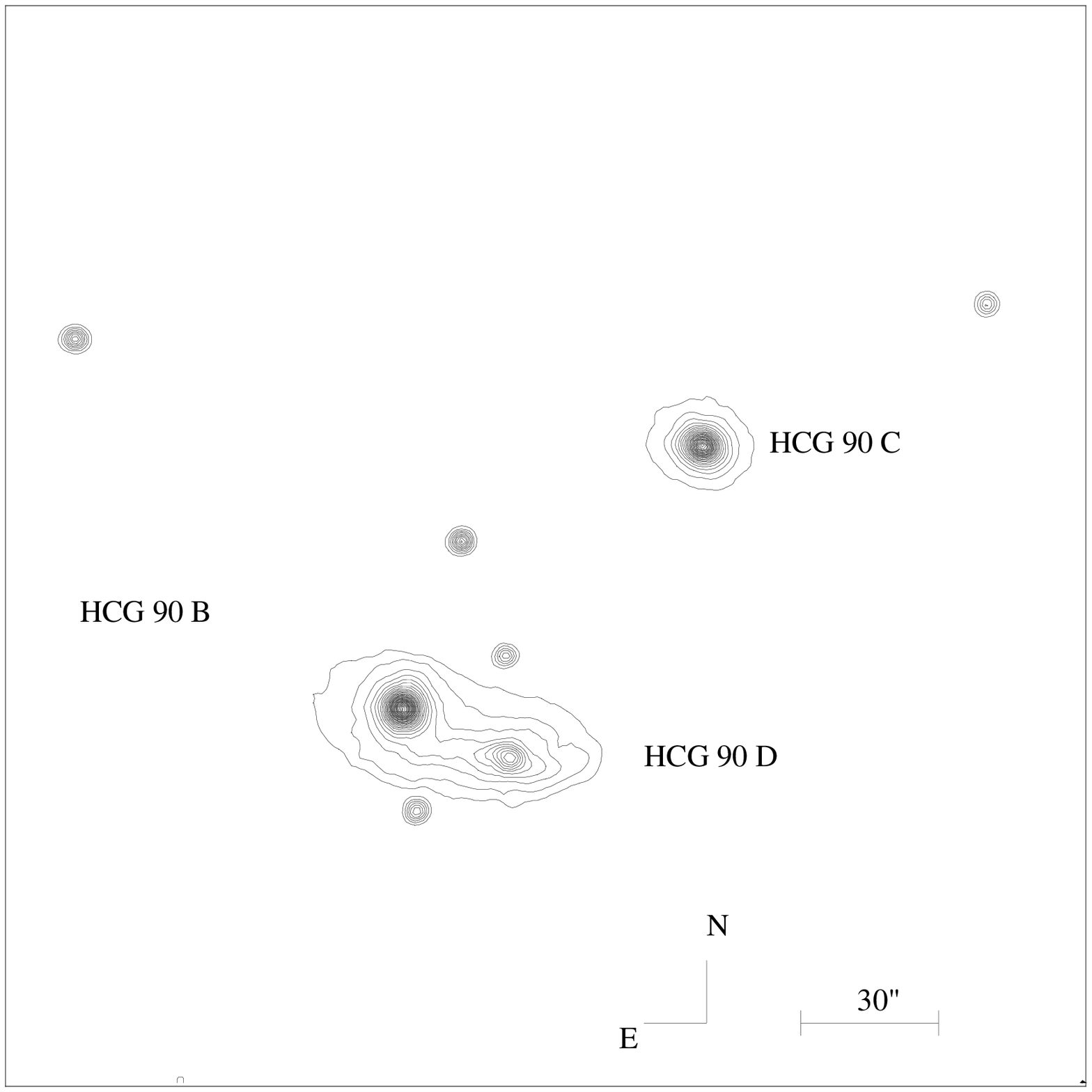]{Continuum image of H90b, c and d. 
The contours (in arbitrary units) were plotted
after a rectangular smoothing with a box of 3 $\times$ 3 
pixels.\label{fig1}}

%
%Fig. 2
%

\figcaption[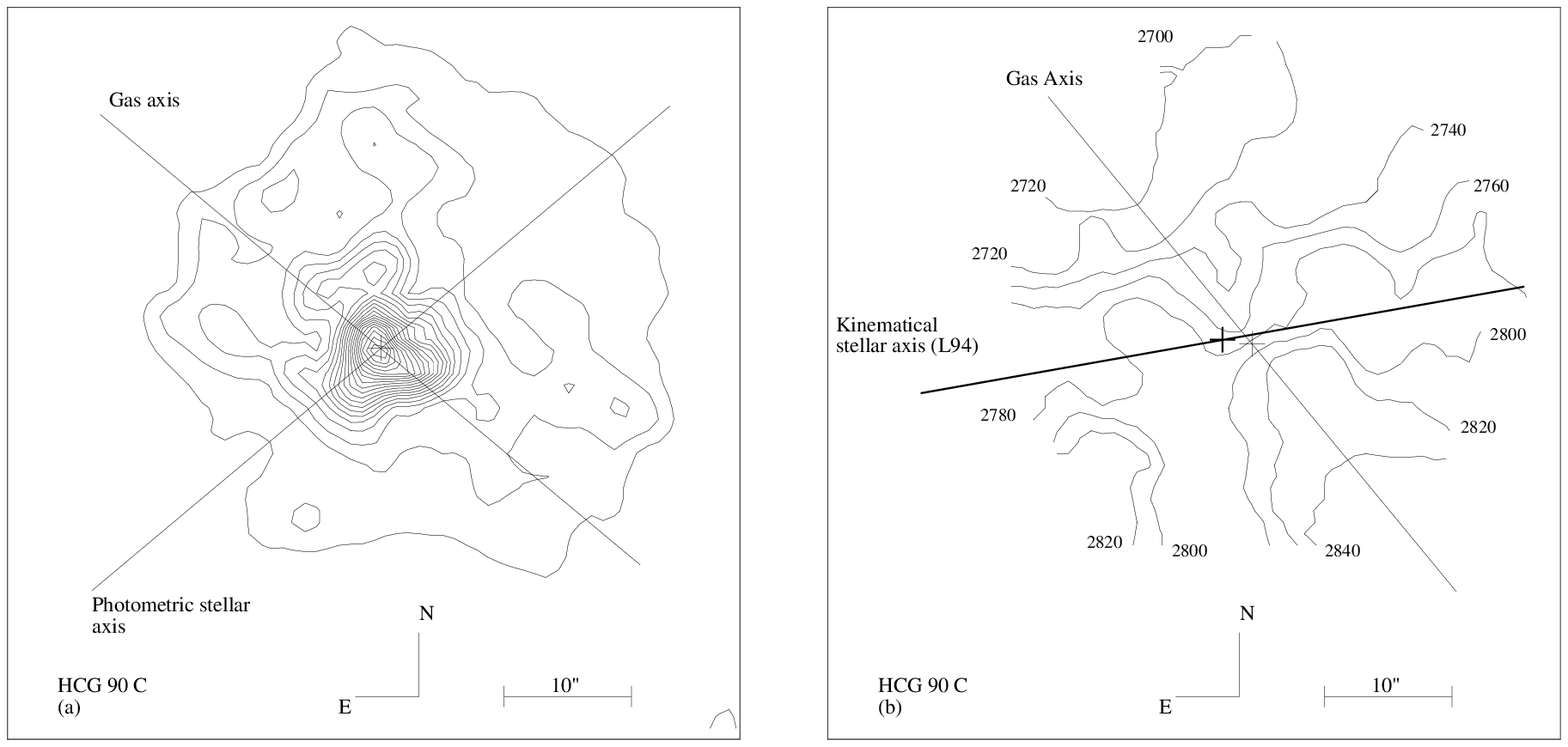]{(a) 
Monochromatic image of H90c in the H$\alpha$ line.
The iso-intensities are
displayed in 10$^{-18}$ erg s$^{-1}$ cm$^{-2}$ arsec$^{-2}$ units. The 
lowest level is 2.4 and the step is 3.5. The position of the stellar 
major axis as determined
from the continuum image and the position of the 
kinematic major axis of the gas are labeled. The cross represents
the monochromatic center, which is coincident with the continuum center
within a seeing disk. 
(b) Velocity field of the gas component. 
The positions of the kinematic major axis of the gas 
(determined in this work) and
the stellar kinematic major axis (from L94) are labeled. 
The cross represents the gas kinematic center and the cross in bold face
represents the continuum center of the galaxy.\label{fig2}}

%
%Fig. 3
%

\figcaption[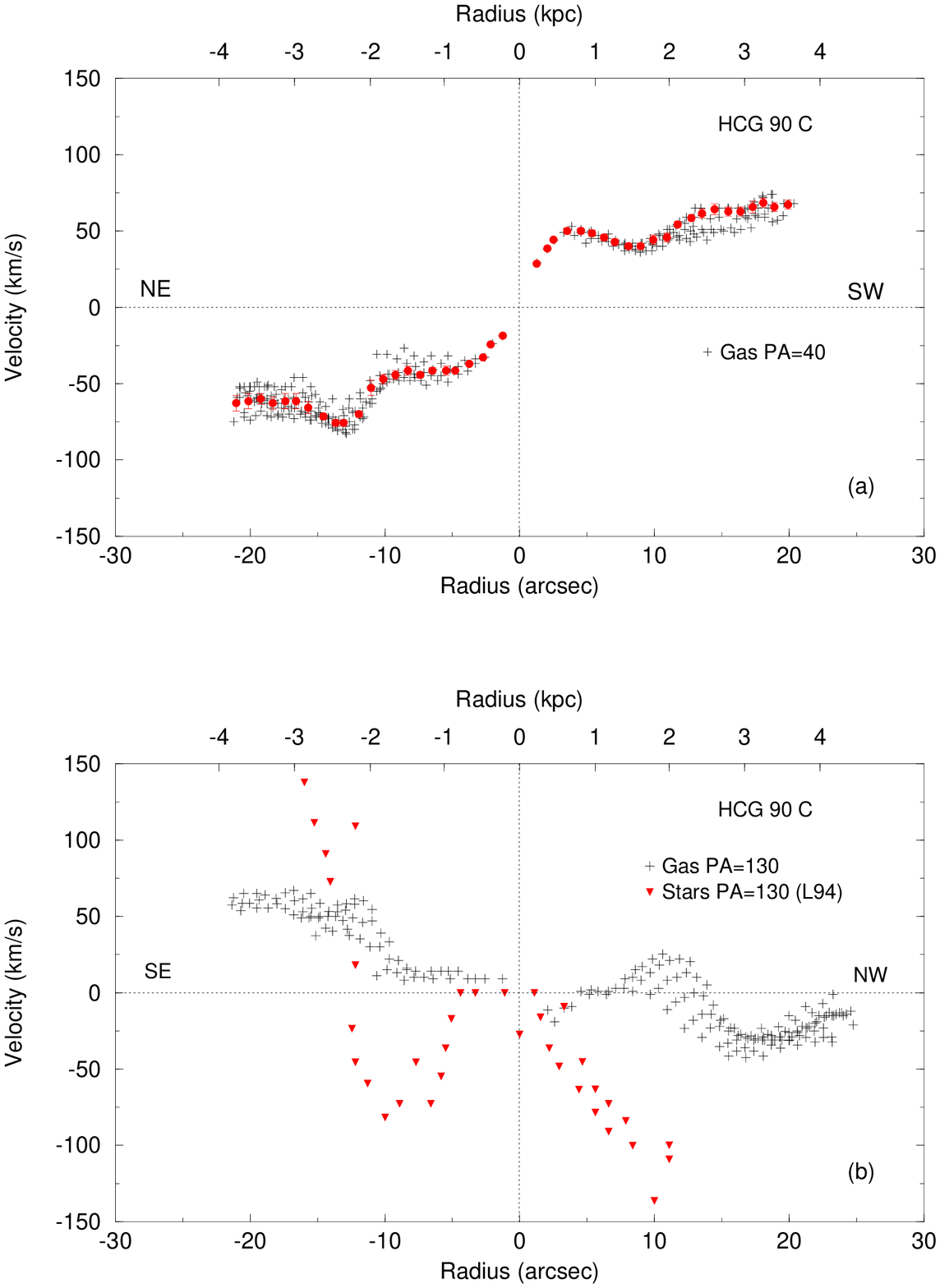]{(a) Average line-of-sight velocity curve for H90c 
(filled circles), 
at PA=40$\degr$ (i.e., the
kinematic major axis for the gas component of this galaxy).  The
horizontal dotted line corresponds to a central gas velocity of 2785 km
s$^{-1}$ (the kinematic center).  
Crosses represent all measured values within 15$\degr$ of the
kinematic major axis (PA=40$\degr$).
(b) VC of the gas component of H90c 
along a PA=130$\degr$ (i.e., the kinematic minor axis of the gas component of
this galaxy) using the continuum center.
Crosses represent all
measured gas velocities within 15$\degr$ of 
PA=130$\degr$, in order to simulate a long slit and allow a comparison
with the stellar VC.
Closed triangles represent the
stellar VC of H90c (from L94)
along a slit with the same PA. The horizontal line marks the
velocity at the continuum center, 2771 km s$^{-1}$. \label{fig3}}

%
%Fig. 4
%

\figcaption[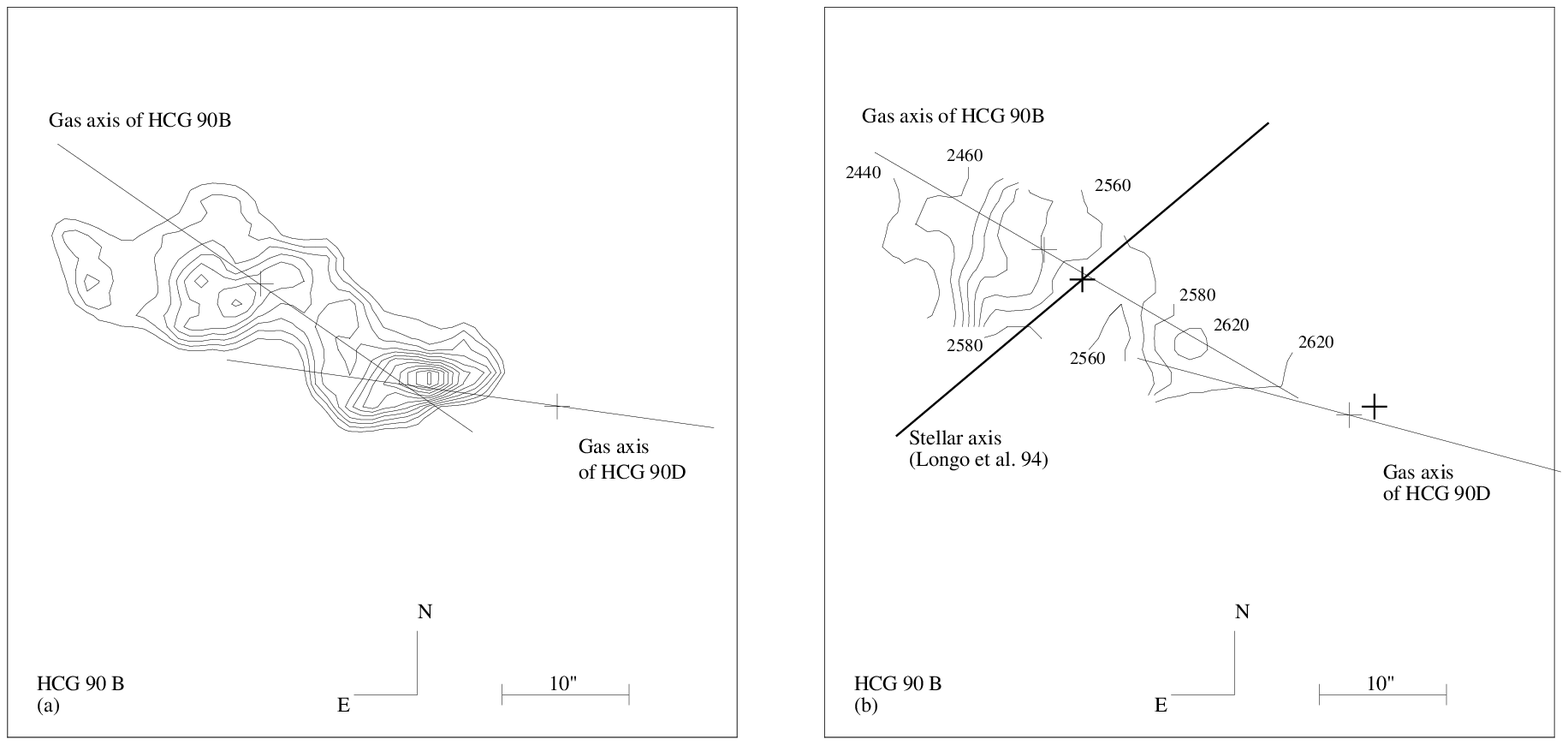]{(a) Monochromatic image of H90B in the H$\alpha$ line. 
The 
iso-intensities are displayed
in 10$^{-18}$ erg s$^{-1}$ cm$^{-2}$ arsec$^{-2}$ units. The
lowest level is 8.9
and the step is 4.3. Crosses represent 
the continuum center for H90b and H90d.
The positions of the gas major axes for H90b and H90d (as
measured from the H$\alpha$ images) are labeled. The
stellar
photometric major axis (measured from the continuum image) 
is almost coincident with the gas major axis.
(b) Velocity field for the gas component of H90b. Crosses represent the gas 
kinematic centers of H90b and H90d. The crosses in
bold face are the corresponding continuum centers. The kinematic major
axes for the gas components are labeled for both galaxies. Also shown is
the kinematic major axis for the stellar component (from L94). \label{fig4}}

%
%Fig. 5
%

\figcaption[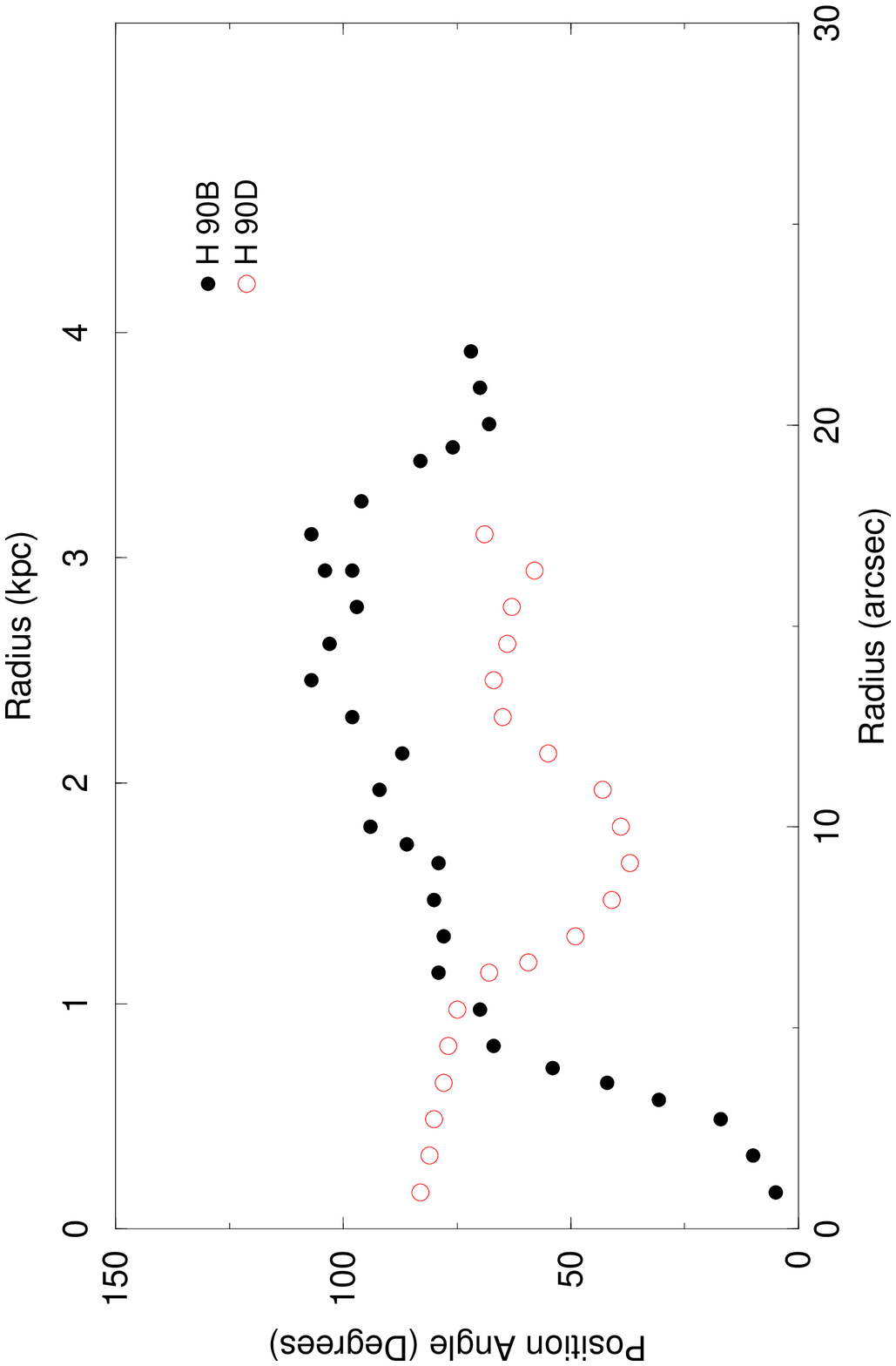]{Variation of the 
PA of the major axis as a function of distance to the 
center of the galaxy (Radius) for galaxies H90b and d, as
measured from the geometry of their gas velocity fields. \label{fig5}}

%
%Fig. 6
%

\figcaption[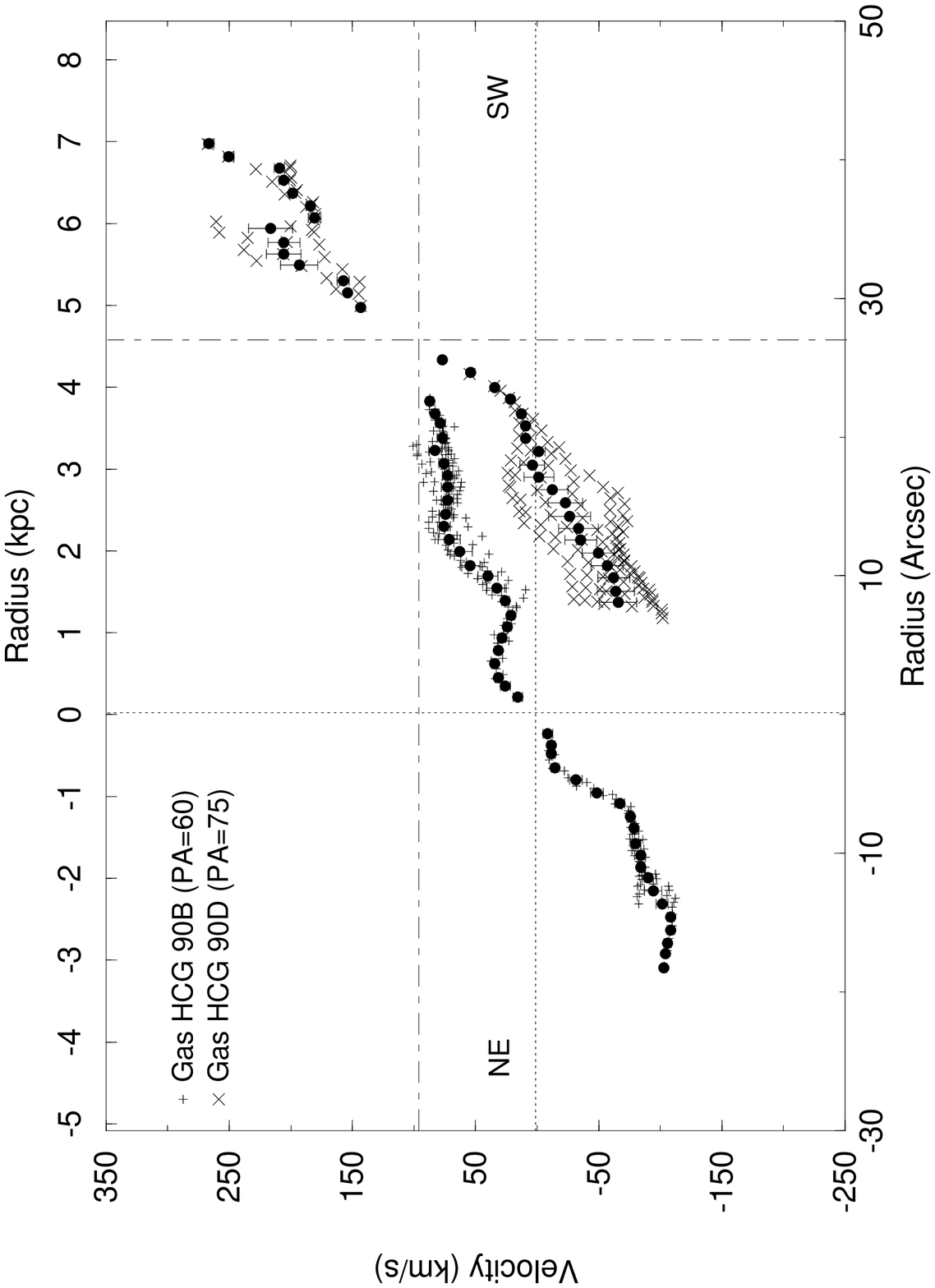]{Line-of-sight velocity curve (VC) of H90b and d.  Plots 
have been made
at PA=60$\degr$ for H90b and at PA=75$\degr$ for H90d (these correspond
to the major axes defined by the geometry of the 
gas kinematics of these galaxies). 
The vertical
long-short line represents the kinematical center of H90d (located 27$\arcsec$
to the SW of H90b). The horizontal long-short line represents the
systemic 
velocity of H90d (i.e., 2635 km s$^{-1}$). The vertical dotted line is
the center of H90b and the horizontal dotted line is the systemic velocity
of H90b (i.e., 2540 km s$^{-1}$). The symbols "$\times$" and "$+$") represent all
mesured values within 15$\degr$ of the kinematic major axes .
The filled
circles are the average velocities at each radius. \label{fig6}}

%
%Fig. 7
%

\figcaption[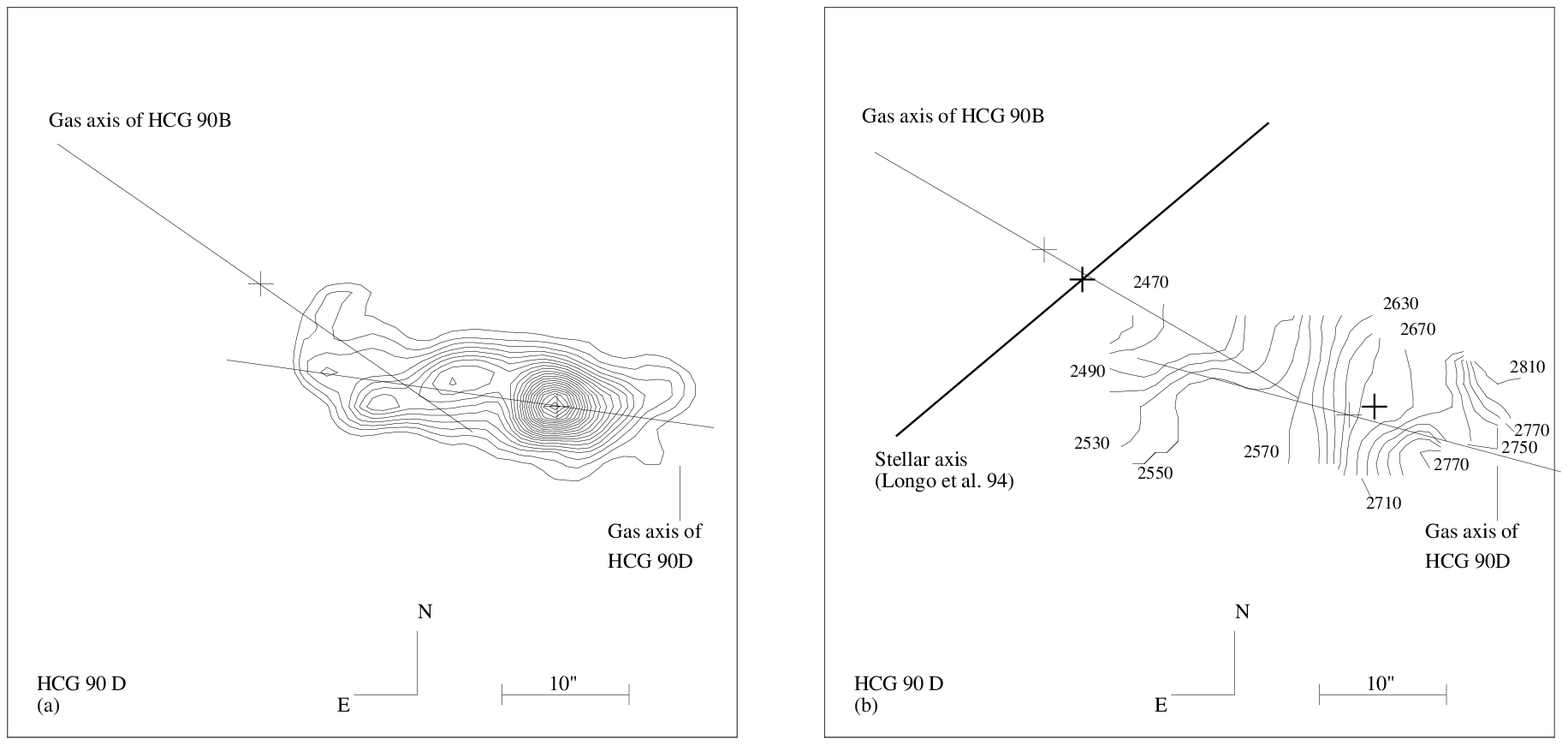]{(a) Monochromatic image of H90d in the H$\alpha$ line. 
The iso-intensities are
displayed 
in 10$^{-17}$ erg s$^{-1}$ cm$^{-2}$ arsec$^{-2}$ units. The lowest level 
is 3.0. The step is 1.2. The crosses represent the monochromatic 
centers of H90b and H90d. These are coincident with
the positions of the continuum centers. 
The gas major axes for the two galaxies are labeled.
The stellar photometric major axis is not shown because it
is very close to the gas major axis.
(b) Velocity field for the gas component of H90d. 
The crosses represent the gas 
kinematic centers of the two galaxies. 
The crosses in bold face represent the continuum centers. 
The kinematic gas major
axes are labeled. Also shown is the kinematic major axis of the stellar
component (from L94). \label{fig7}}

%
%Fig. 8
%

\figcaption[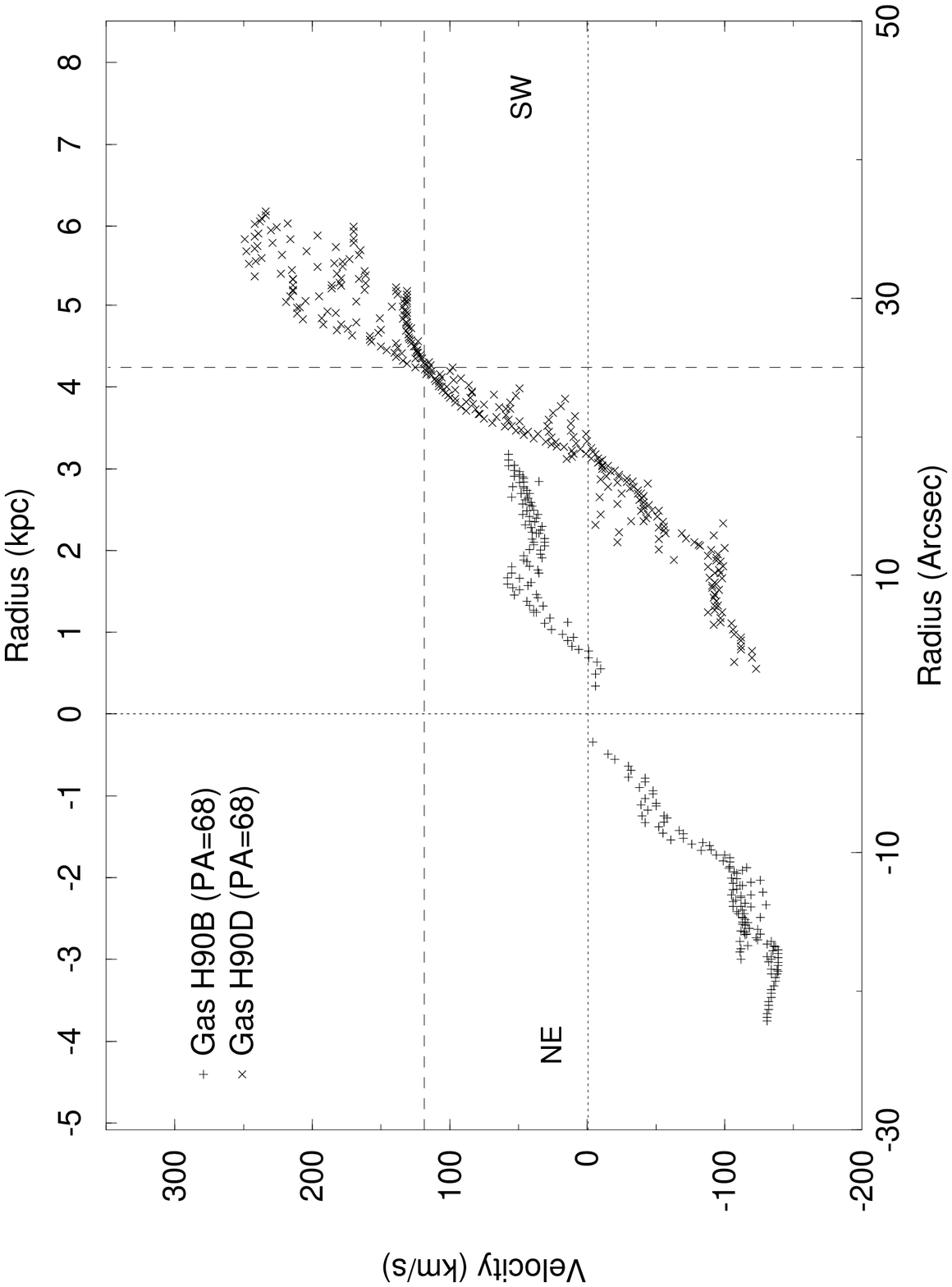]{Line-of-sight velocity curves of galaxies H90b and d 
along
a PA=68$\degr$. The curves have
been obtained by doing a cross section of 
the velocity maps of the galaxies at PA=68$\degr$ and
using the continuum center of H90b as a reference point (position
0,0 on this plot), 
in order to be consistent with L94 and allow a comparison with their
Fig. 7a.
The dotted horizontal line is the velocity of H90b 
at its continuum center 
(i.e., 2570 km s$^{-1}$) and the horizontal dashed line is the velocity at
the continuum center of H90d (i.e., 2688 km s$^{-1}$).
The dotted and dashed vertical lines
represent the centers of H90b and H90d respectively. 
The symbols "$\times$" and "$+$" represent all measured values
within 15$\degr$ of the axis with PA=68$\degr$, in 
an attempt to mimic a long-slit
at this position angle. \label{fig8}}

%
%Fig. 9
%

\figcaption[]{Profiles in the overlaping region of H90b+d. Upper panel
shows the continuum map in the region between H90b and H90d. H90b is in
the NE and H90d is in the SW of the map (north is up and east is to
the left). The bold face square represents the
area where the profiles have been extracted. The lower left panel 
shows the data.
Each square represent a pixel, or 0.91$\arcsec$ on the sky, and an interval of
380  km s$^{-1}$ in velocity space. The lower right panel 
shows a gaussian fit 
to the profiles (FWHM of 63  km s$^{-1}$). \label{fig9}}

%
%Fig. 10
%

\figcaption[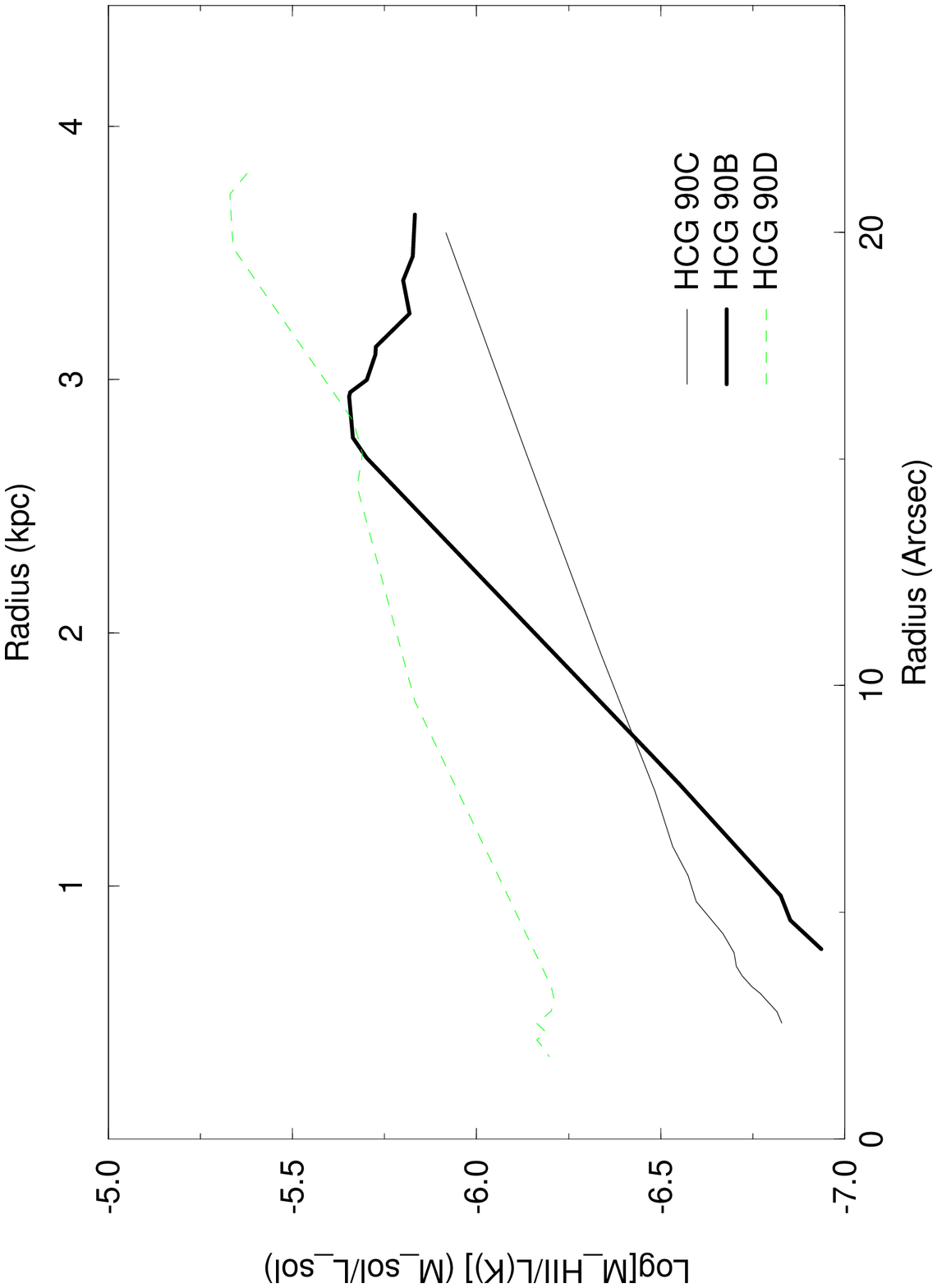]{
The ratio of the ionized gas mass to K band luminosity
as a function of radius for H90b, H90c and H90d. 
The absolute scale for the K photometry is not accurate but
the relative differences in the shapes and intensities of these 
curves are not affected by this uncertainty.
\label{fig10}}

\plotone{plana.fig1.ps}

\centerline{Fig. 1}

\epsscale{1.1}
\plotone{plana.fig2.ps}

\centerline{Fig. 2}

\epsscale{0.85}
\plotone{plana.fig3.ps}

\centerline{Fig. 3}

\epsscale{1.1}
\plotone{plana.fig4.ps}

\centerline{Fig. 4}

\epsscale{0.9}
\plotone{plana.fig5.ps}

\centerline{Fig. 5}

\epsscale{0.9}
\plotone{plana.fig6.ps}

\centerline{Fig. 6}

\epsscale{1.1}
\plotone{plana.fig7.ps}

\centerline{Fig. 7}

\epsscale{0.9}
\plotone{plana.fig8.ps}

\centerline{Fig. 8}

\epsscale{0.85}
\plotone{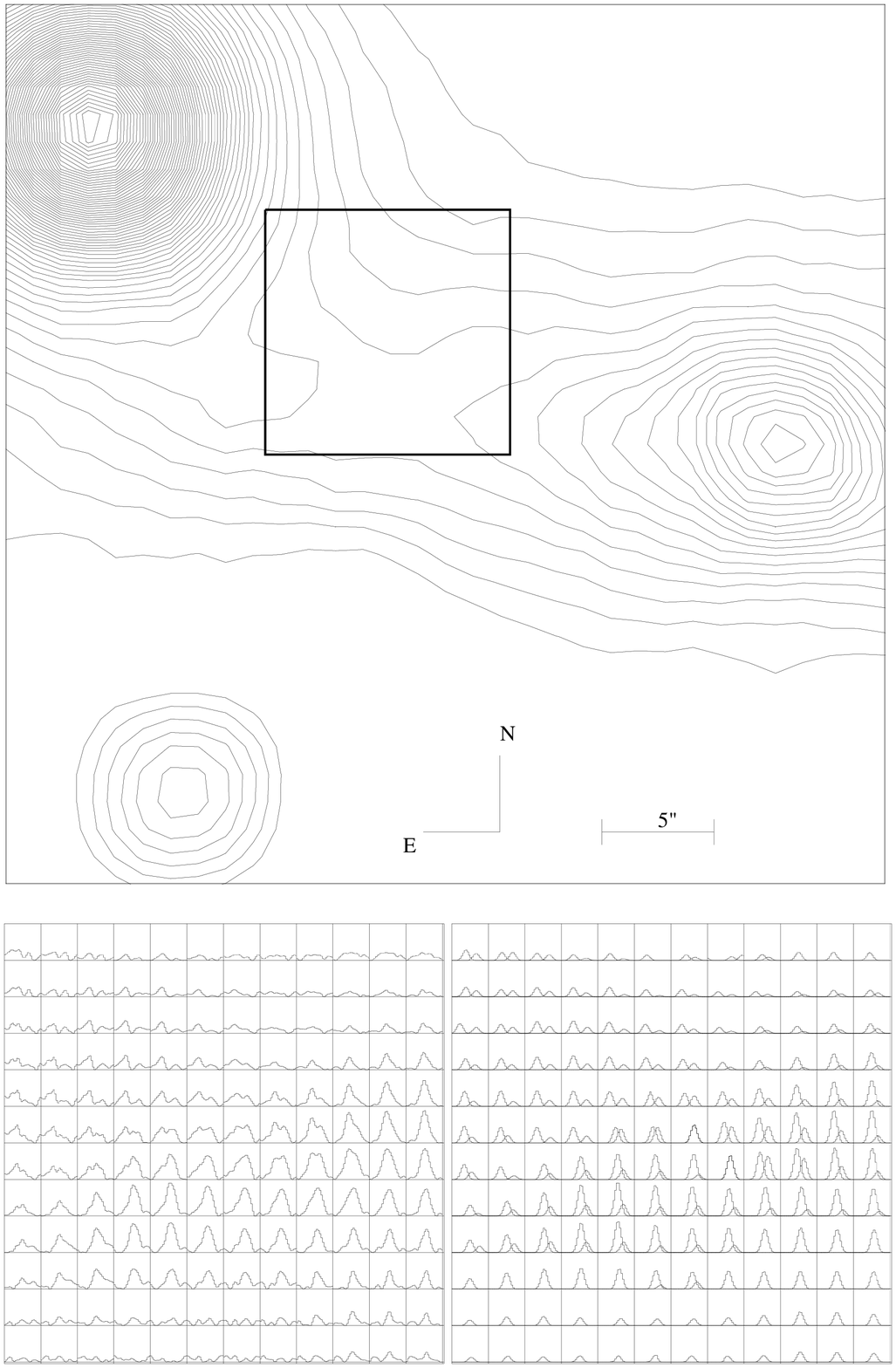}

\centerline{Fig. 9}

\epsscale{0.85}
\plotone{plana.fig10.ps}

\centerline{Fig. 10}

\end{document}